\documentclass[conference]{IEEEtran}
\IEEEoverridecommandlockouts
\usepackage{cite}
\usepackage{amsmath,amssymb,amsfonts}
\usepackage{algorithmic}
\usepackage{graphicx}
\usepackage{textcomp}
\usepackage{xcolor}
\def\BibTeX{{\rm B\kern-.05em{\sc i\kern-.025em b}\kern-.08em
    T\kern-.1667em\lower.7ex\hbox{E}\kern-.125emX}}
\begin{document}

\title{A Low-overhead Kernel Object Monitoring Approach for Virtual Machine Introspection}

\author{\IEEEauthorblockN{Dongyang Zhan\IEEEauthorrefmark{1}, Huhua Li\IEEEauthorrefmark{1}, Lin Ye\IEEEauthorrefmark{1}, Hongli Zhang\IEEEauthorrefmark{1}, Binxing Fang\IEEEauthorrefmark{1}\IEEEauthorrefmark{2} and Xiaojiang Du\IEEEauthorrefmark{3}}
\IEEEauthorblockA{\IEEEauthorrefmark{1}Harbin Institute of Technology, Harbin, 150001, China\\
Email: \{zhandy, 1140310223, hityelin, zhanghongli, bxfang\}@hit.edu.cn\\
\IEEEauthorrefmark{2}Institute of Electronic and Information Engineering of UESTC in Guangdong, Dongguan, 523808, China\\
\IEEEauthorrefmark{3} Department of Computer and Information Sciences, Temple University, Philadelphia, PA, USA\\
Email:dxj@ieee.org
}}

\maketitle
\renewcommand{\thefootnote}{}
\footnotetext{Corresponding author: Lin Ye (Email: hityelin@hit.edu.cn)}

\begin{abstract}
Monitoring kernel object modification of virtual machine is widely used by virtual-machine-introspection-based security monitors to protect virtual machines in cloud computing, such as monitoring dentry objects to intercept file operations, etc. However, most of the current virtual machine monitors, such as KVM and Xen, only support page-level monitoring, because the Intel EPT technology can only monitor page privilege. If the out-of-virtual-machine security tools want to monitor some kernel objects, they need to intercept the operation of the whole memory page. Since there are some other objects stored in the monitored pages, the modification of them will also trigger the monitor. Therefore, page-level memory monitor usually introduces overhead to related kernel services of the target virtual machine. In this paper, we propose a low-overhead kernel object monitoring approach to reduce the overhead caused by page-level monitor. The core idea is to migrate the target kernel objects to a protected memory area and then to monitor the corresponding new memory pages. Since the new pages only contain the kernel objects to be monitored, other kernel objects will not trigger our monitor. Therefore, our monitor will not introduce runtime overhead to the related kernel service. The experimental results show that our system can monitor target kernel objects effectively only with very low overhead.

\end{abstract}


\section{Introduction}
Monitoring kernel objects is widely used to intercept kernel events to protect computer security dynamically. For instance, monitoring the modification of dentry or inode objects (Linux virtual file system cache object) can intercept file operations since file operations will change the target file's dentry or inode object. The modification of a process' file descriptor table indicates that the process opens or closes files. Monitoring socket objects can intercept network events.

With the development of cloud computing and virtualization, more and more operating systems are running inside virtual machines (VM) of clouds and data centers. Virtualization provides a new architecture to monitoring operating systems running in VMs. Some security problems in ubiquitous computing system as discussed by~\cite{xiao2007internet,du2009transactions,du2008security} can be mitigated by this convenient deployment of virtualized cloud instances. In virtualization architecture, Virtual Machine Monitor (VMM) provides virtualization service for VMs and has the highest privilege. It isolates different VMs and is strongly isolated from VMs. Therefore, security monitors running in VMM instead of guest VMs are more secure and transparent. This monitor technology is called Virtual Machine Introspection (VMI)\cite{garfinkel2003virtual}.

By using VMI technology, many security systems\cite{zhan2016cfwatcher,zhan2017protecting,hizver2014real} are proposed to monitor the modification of kernel objects in order to monitor VMs. CFWatcher\cite{zhan2016cfwatcher} intercepts dentry and inode object modification to protect VM filesystem in real time. When the target file's dentry or inode object is modified, the corresponding file operation can be intercepted by the monitor running inside VMM. RTKDSM\cite{hizver2014real} monitors many kernel objects (such as FILE\_OBJECT) to achieve VMI. However, these out-of-VM monitors leverage Intel VT or Intel EPT technologies to perform page-level monitoring, which may introduce overhead to the target VM. Since other kernel objects are also stored in the monitored memory pages, the accesses to them will also trigger the monitor. The false triggering improves the overhead of these tools. Therefore, how to reduce the overhead of these page-level monitors is important.

In this paper, a low-overhead kernel object monitor approach is proposed to reduce the overhead of page-level VMI-based kernel object monitors. The core idea is to migrate the target kernel objects to a protected memory area. As a result, the accesses to other kernel objects will not cause false triggering. Our system is an agentless approach running inside VMM and the privileged VM so it can be used to enhance the VMI monitors. Our target is kernel objects, so we do not need to modify the target VM's kernel. Our system is automated, which means the allocation of the protected memory area and the object migration are performed automatically.

To build protected memory area automatically, we leverage system call reuse technique to let the target VM build it. After that, we analyze the target VM's memory to detect the target object and the related pointers. An automated pointer detection method is proposed to detect the target object's related pointers automatically. Then, the target object is migrated to the protected memory area	 and the related pointers are updated to point to the new migrated object. Finally, the protected memory area is monitored and protected by using hardware-based events.

In summary, our paper has several contributions, which are as follows.
\begin{itemize}
  \item A low-overhead monitoring approach is proposed to reduce the overhead of VMI-based page-level kernel object monitors, which migrates target kernel objects to a protected memory area.
  \item Protected memory areas can be created automatically by leveraging system call reuse technique. An automated kernel object migration method is proposed.
  \item The experimental results show that our system can effectively reduce the runtime overhead of VMI-based monitors.
\end{itemize}

The rest of this paper is organized as follows. Section \ref{S:related-work} summarizes the related work. The system design and implementation are described in Section \ref{S:design}. Section \ref{S:evaluation} evaluates the effectiveness and performance of our system. The conclusion is given in Section \ref{S:conclusion}.

\section{Related Work}\label{S:related-work}
During the development of cloud computing, security is always a hot topic. Some related security issues are studied in~\cite{du2006adaptive,du2007effective,xiao2007survey,du2005designing}. To protect VMs, Virtual Machine Introspection \cite{garfinkel2003virtual} technology is proposed. It is an out-of-the-box technology to monitor VMs and is widely used in cloud security. In the virtualization architecture, VMM provides virtualization service for VMs , which has the highest privilege and is strongly isolated from the VMs. As a result, the security monitor running inside the VMM can access the VM hardware execution state. Based on VMI technology, many security monitor systems \cite{zeng2015pemu,shi2018vanguard,dolan2013tappan} have been proposed to protect VMs. Compared with the in-VM security tools \cite{adukkathayar2015advanced,kaczmarek2014operating}, VMI-based monitors are more secure and transparent.

Many VMI-based security monitors\cite{zhan2016cfwatcher,zhan2017protecting,hizver2014real} enhance the VM security by intercepting the modification of VM kernel objects. CFWatcher\cite{zhan2016cfwatcher} is a real-time filesystem monitor which can intercept every access to the target files by monitoring the corresponding dentry or inode objects. RTKDSM\cite{hizver2014real} enhances VM security by monitoring many kernel objects (such as FILE\_OBJECT) from outside. Because of the page-level intercepting, the extra overhead is introduced.

The semantic gap is a major challenge for VMI-based monitors. Monitors running inside the VMM can only obtain the hardware binary execution information, but they need the high-level execution information (such as system calls) to perform introspection. The gap between binary execution information and high-level information is called semantic gap. To narrow the semantic gap, many works have been proposed in recent years. Virtuoso\cite{dolan2011virtuoso} reuses VM code to generate inspection tools in VMM automatically. Firstly, it traces the execution of programs running inside VMs to get some necessary code pieces. Then, it modifies and reuses them to generate the inspection tools, which can run in the VMM. SYRINGE\cite{carbone2012secure} reuses some functions inside the target VM. A break point is injected in the target VM. When this point is triggered, SYRINGE suspends the normal VM execution and then lets the VM execute some specific reuse functions to bridge the semantic gap. In addition, it can verify the integrity of the reuse functions. ShadowContext\cite{wu2014system} leverages system calls redirection technology to bridge sematic gap. By converting a process within the target VM to a dummy process, ShadowContext leverages the dummy process to invoke some system calls to narrow the semantic gap automatically. SAVM\cite{zhansavm} and HyperShell\cite{fu2014hypershell} are some in-VM management tools, which also leverage the system call reuse technology.

To provide security service for cloud users, many systems have been proposed. CloudVMI\cite{16wook2014cloudvmi} provides the VMI interface for cloud customers, which can help them enhance the VM security. Furnace\cite{bushouse2018furnace} enables cloud customers to execute their VMI-based code in the VMM with high performance. Several papers (e.g., \cite{xiao2007internet,du2009transactions,xiao2007survey,du2007effective,du2008security})have studied related issues.

In this paper, we aim to reduce the runtime overhead of VMI-based kernel object monitors, which can be used by many VMI-based security monitors.

\section{Design \& Implementation}\label{S:design}
\subsection{System Overview}
The traditional approach is to directly monitor the memory pages of the target kernel objects. Therefore, the accesses to other kernel objects will cause false triggering and unnecessary overhead. To solve this problem, we allocate a protected memory area then migrate target kernel objects automatically.



There are some requirements of our system.

\begin{itemize}
	\item\textbf{Effectiveness.} Firstly, all of the related pointers after the kernel object migration should be modified. It is however, a challenge to find all of the pointers. Searching the entire kernel memory to look them up is a possible way. We can check if the memory data equals to the pointer. But, it is not easy to identify whether the content is a related pointer or just a data. Modifying the wrong memory area may cause VM crash. In addition, the release of original kernel object is another problem. After the migration is completed, it will inevitably cause memory holes if the original kernel object cannot be effectively released.
	
	\item\textbf{Performance.} The object migration should introduce acceptable overhead to the target VM and can highly improve the performance of the traditional VMI-based kernel object monitors.
	
	\item\textbf{Security and transparency.} The monitor should be fully secured and be isolated from the target VM.
	
\end{itemize}

\begin{figure}
	\centering
	\includegraphics[width=0.9\linewidth]{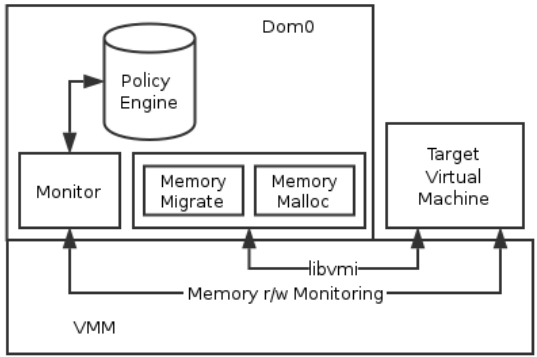}
	\caption{System Architecture}
	\label{fig:s-a}
\end{figure}

In order to meet these requirements, we propose a low-overhead kernel object monitoring approach by using kernel object migration. The overall system architecture is illustrated in Figure \ref{fig:s-a}.

The whole system includes monitor module, memory operation module and policy engine, where memory operation module could be divided into memory migration module and memory allocation module. The whole system is deployed in VMM and the privileged VM, which is completely isolated from the target VM. More details about work flow are as follows:

\textbf{Step 1.} With the help of the VMI-based toolkit LIBVMI and system call reuse technology, memory allocation module allocates protected memory pages in the target VM kernel memory space, then obtains the start address of protected area.

\textbf{Step 2.} The memory migration module copies original objects to a protected memory area. All of the related pointers are retrieved based a proposed policy and re-pointed to the new kernel objects. Finally, the memory space of original kernel objects will be released.

\textbf{Step 3.} The monitor module dynamically monitors all of the accesses to the protected memory pages based on the security engine.

\subsection{Automated Memory Area Allocation}
We leverage system call reuse technology to automatically allocate a kernel memory area. The system call reuse technology is widely used by VMM to perform in-VM management (such as killing processes and reading files). Since most of the operating system's critical operations are performed by system calls, reusing the system calls of the target VM can help VMM to control VM execution from outside automatically. For instance, if we want to kill a process in VM, we can inject the system call KILL with specific parameters to the target VM.

The system call reuse technology usually injects a system call to the target VM through a series of operations such as modifying some registers and memory data when the VM is invoking system call. After the injected system call has been finished, VMM obtains the execution results and restores execution context of the original system call. Then, the VM kernel will execute original system call properly and finally return to user space.

In a Linux operating system, programs can obtain and release the kernel memory space with continuous physical addresses by using kmalloc() and kfree(). By tracking the kmalloc() and kfree() functions, we find that the main system calls involved by them are MMAP and MUNMAP. MMAP is responsible for mapping contents of a file to the memory space of the process. After the mapping, process can read or write the file by reading and writing the memory. By using this system call, we can let the target VM allocate a memory space through passing the specified memory address and length. This memory can be used to store the migrated kernel data. MUNMAP can release the memory mapping created by MMAP, which contains two parameters: 'start' (the head address of memory to release) and 'length' (the length of memory to release).

\begin{figure}
	\centering
	\includegraphics[width=\linewidth]{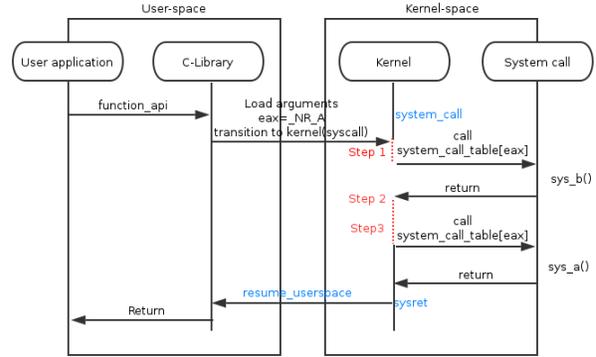}
	\caption{System Call Injection}
	\label{fig:syscall-injection}
\end{figure}

The core idea of automatically allocating protected memory area is to use system call reuse to inject the MMAP into the target VM and then get the execution results, which steps are shown in Figure \ref{fig:syscall-injection}.

\textbf{Step 1: Replacing the original system call.} We replace the first instruction of the system call entrance with INT3 instruction by re-writing the target VM memory. When processes invoke system calls, the target VM will reach the system call entrance. Then, INT3 instruction will trigger a VMExit event and the control will be transferred to VMM. After saving the context, including related memory and register content, we replace the register content and related memory content with the parameters of the injected system call. Finally, the target VM is resumed and executes the injected system call.

\textbf{Step 2: Obtaining the injected system call results.} Similarly, we also replace the system call exit instruction with INT3 instruction, which can intercept the the return of the injected system call. When the injected system call is finished, the target VM will be suspended. At this time, we can obtain the execution results from VMM, which include related register and memory content.

\textbf{Step 3: Recovering the original system call.} After successfully obtaining the corresponding results, we restore the original system call context by setting the corresponding registers and memory to the previously saved execution information. Then, the EIP is set to the system call entrance. Therefore, the target VM will execute the original system call and return to user space.

With the system call reuse technology, we can complete two tasks: a) Insert a MMAP to get protected memory area starting at specified kernel address. b) Use the MUNMAP to release protected memory pages which are in free. Above all, we can achieve the goal of automated memory area allocation.

\subsection{Automated Object Migration}
The migration process will be described with two use cases. The first one is migrating dentry objects, another is migrating file descriptor table (fdt).

\textbf{Dentry Structure}

Whenever a VM opens a new file, the operating system will create a new dentry object. Every dentry object is placed on the hash table 'dentry\_cache'. With the semantic knowledge of the target VM operating system, we can obtain the address of 'dentry\_cache' in memory. Then, the head address of all dentry objects can be obtained from the whole hash table. If we want to migrate a dentry object, we can copy it to the protected memory area. Since the length of a dentry object is fixed, all of dentry objects can be continuously copied to the protected memory pages.

After copying a dentry object, we need to modify the related pointers. By analyzing dentry object, we can build a profile of this object. The pointers related to dentry object can be divided into three types, which are as follows:

\begin{enumerate}
  \item External pointers to dentry object;
  \item Internal pointers to internal addresses of dentry object;
  \item Internal pointers to the next/previous dentry objects.
\end{enumerate}

We handle these three types separately.

\begin{figure}
	\centering
	\includegraphics[width=\linewidth]{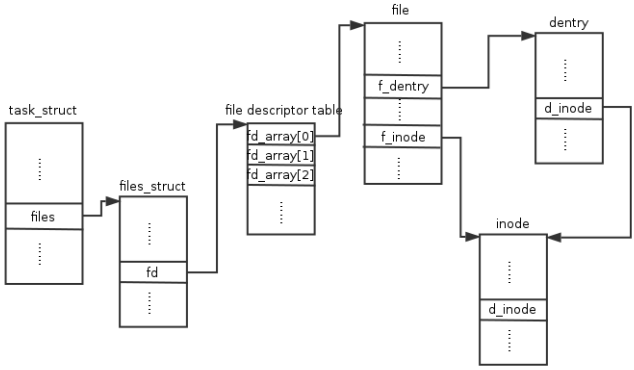}
	\caption{Relationship of Kernel Objects}
	\label{fig:r-o}
\end{figure}

\textbf{Type 1.} There are many external pointers pointing to dentry object. 
For instance, There is a field named 'f\_dentry' in file object (another important kernel object in filesystem), which points to the related dentry object. The file object also has a field named 'f\_inode' pointing to the corresponding inode object. Since related dentry and inode objects have pointers linking each other, we can use these pointers to verify if we get the right pointers. The relationship among inode, dentry, and file is shown in Figure \ref{fig:r-o}.

\textbf{Type 2.} Inside the dentry structure, there is a string pointer named 'd\_name' pointing to an internal memory field which offset is fixed. This memory area stores the file/directory name. After migration, we need to change these pointers to the new addresses. Based on the profile, we are able to calculate and rewrite the new address because the offset is fixed.

\textbf{Type 3.} Because the dentry objects are linked by doubly linked list, we can find the next/previous nodes from the old dentry object. After the migration, we need to change the corresponding fields of them to make sure that they can reach the migrated object.

After completing the dentry migration, we set the reference count of the original dentry object to 0 and add the original dentry object to LRU linked list. Finally, the operating system will eventually release the data structure on the LRU linked list.

\textbf{File Descriptor Table}

Every process in memory maintains a file descriptor table (fdt), which contains many file descriptors(fd) pointing to the opened files. The fdt address of each process can be obtained via task\_struct$\rightarrow$files$\rightarrow$fdt. If we want to migrate this object, it is easy to copy it to protected memory pages and modify corresponding fdt pointers in the process structure. Actually, the file descriptor table is an array of file pointers, which point to every opened file of the process. The related pointer, which points to this structure, is only in the process structure. So, we just need to modify the corresponding pointer after the migration.

\subsection{Memory Operation Monitor}
After the automated kernel object migration, the monitor based on VMI can perform page-level monitoring on the target memory pages. With the help of the Intel EPT technology, we can easily monitor all read/write operations on the protected memory area. In our implementation, LIBVMI is used to help us to perform VM page-level monitoring. LIBVMI is a VMI c-library, which makes it easier to monitor behaviors of VM.

By registering the monitoring events of the target memory pages, it will trigger a memory event when VM operation system reads or writes the corresponding kernel objects. The listener outside VM can obtain the details of the event through the callback function then analyze the behaviors based on the policy engine.

\subsection{Analysis \& Discussion}
Our system is secure, because it runs in VMM. Compared with traditional approaches of directly monitoring the memory pages, which include target kernel objects, our approach has a lower overhead and better performance.

To further ensure the migration security, cloning the target VM is another possible way to test the effectiveness. Before the migration, we can clone the target VM and perform the migration. Then, we can let the cloned VM access these files to check if there are some errors. If no error is triggered, we can perform the migration in the target VM. To let VM access files, we can also leverage system call reuse technology, which injects OPEN and READ system calls into the cloned VM.

Kernel object monitoring faces some security challenges. A sophisticated attacker may firstly get the root privilege of the target VM, and then he could also migrate the target kernel object to another kernel space. This attack may make the kernel object monitor useless, because the monitor only monitors the target object's memory area, not including the related pointers. But, this challenge can be addressed by many ways. In most of cases, the attackers have to read the old object to migrate it to other place, which will trigger the monitor. For instance, the attacker needs to firstly get the content of the target file's dentry and then copy it to another kernel space, which will trigger the read monitoring. Monitoring the virtual file /dev/mem is another possible way, because most of kernel modifications are performed by using this file. 

Monitoring the lifecycle of the target kernel object is necessary for kernel-object-based monitors. Related work\cite{zhan2016cfwatcher} has already proposed the solve method, which keeps the object alive in the VM kernel.

\section{Evaluation}\label{S:evaluation}
After implementation, we evaluate the prototype system to test the the effectiveness, performance and overhead. We use a 3.4GHz Intel i7 CPU computer with 8GB memory as the testbed, supporting Intel VT. Xen is selected as the VMM, which is a widely used hypervisor. The operating systems of Dom0 and the target VM are Ubuntu 16.04 x64 and Ubuntu 12.04 x64 respectively. The target VM is configured with 1 VCPU and 512 MB memory.

\subsection{Effectiveness}
\textbf{Dentry Object}

There are two test programs named 'dentry\_test\_1' and 'dentry\_test\_2' inside the target VM. Both two programs have the same task: opening the file 'test.txt' in write-only mode, then waiting for the user to type a single character.

We run the test program 'dentry\_test\_1' in the target VM, then we pause the VM when it is waiting for a user to type in. At this time, we allocate the protected memory area in the target VM, then the memory migration module starts to migrate the dentry object of 'test.txt'. It obtains the corresponding dentry object, and the reference count is 1, which means this file is accessed by one process now. Then, this object is  migrated to the protected memory pages, and the related pointers are redirected. Finally, the target VM is resumed, and the memory monitor starts.

After the migration, we run 'dentry\_test\_2'. This program opens the same file and waits for a user to type in. When the file is accessed, our monitor is triggered, which means the program is accessing the migrated object. After resuming the VM, we find that the monitored dentry object is referenced by two programs, and the reference number is two. This result can prove the correctness and effectiveness of the dentry object migration.

\textbf{File Descriptor Table}

We build a test program named 'fdt\_test' in the target VM. The test program opens a file called 'fdt\_1.txt', then it waits for a user to type a characters and opens another file called  'fdt\_2.txt'.

When the target VM has opened 'fdt\_1.txt' and is waiting for user, we start to migrate the fdt to the protected memory pages. The steps are similar with those of dentry. At this time, the fdt has four fd entries, among which the fd of 'fdt\_1.txt' is 3. After resuming the target VM and starting the memory monitor, the program opens a new file 'fdt\_2.txt'. This operation triggers our memory monitor, and the new entry of 'fdt\_2.txt' in fdt is 4. From this result, we can know the migration of fdt is successful.

\subsection{Performance}
In the performance evaluation, we want to test the performance impact of real-time monitoring on the target VM. Since monitoring dentry objects will introduce high overhead to the target VM's filesystem, we test our system's performance with monitoring dentry objects. By using the benchmark of extracting the compressed file of Linux kernel, we evaluate the target VM's filesystem performance when monitoring different numbers of dentry objects before and after object migration. We compare our system with CFWatcher\cite{zhan2016cfwatcher}, which monitors dentry objects to intercept file access. The entire test can be divided into the following three steps:

\textbf{Step 1:} Turn off page monitoring and test the target VM's filesystem performance by extracting the compressed file.

\textbf{Step 2:} Select different numbers of kernel objects, then perform memory page read/write monitoring on the memory pages where the kernel object is located. At this time, we test the target VM's filesystem performance. This is actually what CFWatcher does.

\textbf{Step 3:} All of the kernel objects in Step 2 are migrated to the newly created and protected memory area. Then, the protected memory area is monitored for read and write operations. After that, we test the performance of the target VM to compare with CFWatcher.

\begin{figure}
 	\centering
 	\includegraphics[width=0.85\linewidth]{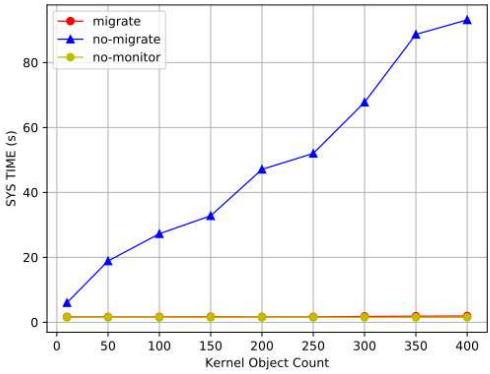}
 	\caption{Relationship between Extracting Time and Object Number}
 	\label{fig:r-t}
\end{figure}

We choose the kernel time of extracting compressed file as the benchmark of read/write performance of the target VM. For Step 1, we test 10 times and compute the average value, which is used as the baseline data. Then, we select and monitor different magnitudes of dentry objects (10, 50, 100, 150, 200, 250, 300, 350 and 400) to test the performance with dentry monitoring. For each object count, we first test the performance without migration (Step 2), and then test the performance with migration (Step 3). Each step is tested for 10 times and the average is calculated, which results are shown in Figure \ref{fig:r-t}.

\begin{figure}
	\centering
	\includegraphics[width=0.85\linewidth]{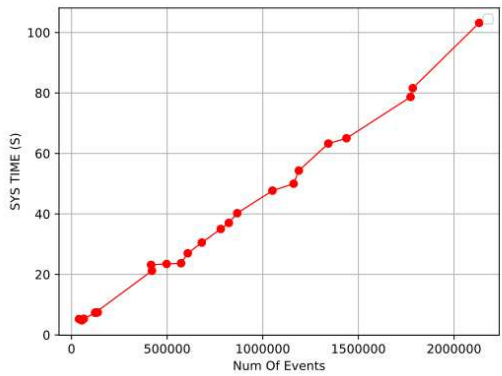}
	\caption{Relationship between Extracting Time and Event Number}
	\label{fig:r-v}
\end{figure}

From the results, we can find that when monitoring a large number of dentry objects, the overhead of non-migration method is high. In contrast, the overhead of our system is very low. But we can find that our system still introduces overhead to the target VM, that is because some of the monitored objects are accessed by other processes. If the selected objects are not touched by any other processes, the runtime overhead will be nearly zero. Actually, this overhead is related to the number of VMExit events triggered by the target VM, which relationship is shown in Figure \ref{fig:r-v}. Our system can significantly reduce the number of VMExit events.

\subsection{Overhead}
The main overhead of our system is extra memory occupation. Since the monitored objects need to be migrated to the protected memory area, this memory area will occupy extra memory in the target VM. In our design, the migrated objects are continuously distributed, so the memory needs not to be very large. In the experiment, when monitoring 400 dentry objects, we leverage 32 pages as the protected memory area, which size is 128 KB.

\section{Conclusion}\label{S:conclusion}
This paper proposes a low-overhead VM kernel object monitoring method for VMI, which migrates the target kernel objects to the isolated memory area to reduce the runtime overhead. By leveraging system call reuse technology, the protected kernel memory area is built automatically. Then, we migrate the target kernel objects by leveraging automated pointer redirection technology. Finally, these memory pages are monitored to intercept memory accesses. The experimental results prove our system's effectiveness and performance. In the future, we are going to apply this approach to monitor more kinds of kernel objects.

\section*{Acknowledgment}
This work was partially supported by DongGuan Innovative Research Team Program under Grants NO. 201636000100038, and National Natural Science Foundation of China under Grants NO. 61872111.

\bibliographystyle{IEEEtran}
\bibliography{sigproc,add}

\end{document}